\documentclass[a4paper]{jpconf}
\usepackage{iopams}
\usepackage{lipsum,babel}
\usepackage{graphics}
\usepackage{color}
\usepackage[usenames,dvipsnames]{xcolor}
\usepackage{dcolumn}
\usepackage{bm}
\usepackage[latin1]{inputenc}
\usepackage{textcomp}
\usepackage{footnote}
\usepackage{epstopdf}
\usepackage{url}
\usepackage{amssymb}
\usepackage{balance}
\usepackage{cite}
\usepackage{hyperref}
\usepackage{amsmath}
\usepackage{gensymb}
\newcommand{\red}{\textcolor{black}}

\def \MgB2 {MgB$_{2}$ }
\begin{document}

\title{\red{Computational Modelling of Russia's First 2G-HTS Triaxial Cable}}

\author{M. Clegg, M. U. Fareed, M. Kapolka \& H. S. Ruiz}
\address{College of Science and Engineering \& Space Park Leicester, University of Leicester, Leicester LE1 7RH, United Kingdom}
\ead{mlc42@leicester.ac.uk ; hsrr1@le.ac.uk} 


\begin{abstract}
A better understanding of the interaction between three phases is required when developing superconducting cables for high voltage AC systems. With a particular focus on the energy losses of real power transmission cables, in this paper we utilize the so-called H-formulation of Maxwell equations to devise a 2D model for superconducting triaxial cables. The major aim of this model is to comprehend and \red{reproduce} the experimental observations reported on the first triaxial prototype cable developed by SuperOx and VNIIKP. The computationally modelled and prototyped cable is made of up to 87 tapes of 4 mm width SuperOx tape arranged across the three phases. Our computational results are compared to the experimental measurements performed by VNIIKP with the electrical contact method, showing a high degree of accuracy over the outer phase of the cable, whilst revealing technical issues with the experimental measurements at the inner phases. Thus, in consultation with VNIIKP it has been concluded that for the actual experimental measurement of the AC losses at the inner phases, and consequently of the overall cable, a sophisticated calorimetric setup must be built. Still our model is capable to \red{provide an independent assessment of the} VNIIKP-SuperOx cable design, by investigating the magnetic profiles per phase in the time domain. In this sense, we confirm that the unbalanced arrange of currents and distancing between the phases affirmatively lead to no magnetic leakages, and therefore to an adequate balance of the cabling inductance. 
\end{abstract}

%
%
%
%

\section{Introduction}\label{Sec.1}

Improving the current density capacity of power transmission is important as the electricity demand across the world continues to increase. Accommodating for this pressure is possible with high temperature superconducting (HTS) cables that allow for high levels of current density while also maintaining a low level of energy losses produced by the system. HTS cables have been investigated over many years as numerous projects have already been installed and tested within several power grids, showing the potential of these cables when applied in this sector \cite{Doukas2019IEEE}.  However, a downfall in the HTS cable production is the capital and operational cost of the HTS wire and refrigeration system \cite{Kalsi2011SC}. Therefore when constructing a HTS cable, further thought needs to be put into the configuration of the cable \red{with the three concentric phases resulting in the most compact design and therefore lowest material consumption \cite{Willen2005CIRED}.} This is especially important when considering a three phased system due to the added energy losses imposed by the alternating currents, which although could be significantly reduced by considering DC currents, depending on the amount of current being transmitted~\cite{Ruiz2018aIEEE,Ruiz2018SUST,Ruiz2013JAP,Ruiz2012APL}, all major transmission applications and nearly all distribution systems operate under AC conditions. Thus, for the development of three-phase networks aided by superconducting cables, one configuration of great interest is the designing of HTS triaxial cables~\red{\cite{Fisher2003IEEE,Fetisov2017IEEE,Sun2019IEEE,Lee2019IEEE,Cao2019E3S,Lee2020Energies,Fetisov2021IEEE}}, capable to offer a high critical current density with a low amount of HTS wire, whilst maintaining the power system balance. 

The possibility of triaxial HTS cables in three-phase power networks has been actively researched since the beginning of this century~\cite{Fisher2003IEEE,Politano2001IEEE}, and since then multiple projects have been conducted such as the Bixby project at Columbus, Ohio~\cite{Demko2007IEEE} and the AmpaCity project at Essen, Germany~\cite{Stemmle2014IEEE} both using the first generation (1G) of Bi-2223 tapes for the manufacturing of their cables. The AmpaCity cable was in fact the first HTS cable to be integrated into a real power system of a major city, and has been the precursor of many major developments of HTS AC cables for power applications~\cite{Malozemoff2015Book}. Nevertheless, more recently the development of triaxial cables has been focused on the use of the so-called second generation (2G) of HTS tapes, with 2G HTS triaxial cables being produced in South Korea \cite{Lee2020Energies} and Russia \cite{Fetisov2017IEEE}. These projects have displayed the capabilities of producing a large scale three phase transmission system using the latest developments on the manufacturing of superconducting tapes, and the possibility to balance their total currents for a null magnetic field outside of the conductor, as the different radii of the three phases produces non-uniform current distributions due to the magnetic induction suffered by the type-II superconductors~\cite{Nassi2000SUST}. 

\begin{table}[!]
\centering
\caption{\label{Table_1} Measured dimensions and critical currents for the triaxial prototyped cable in~\cite{Fetisov2017IEEE}.} 
\begin{tabular}{@{}l*{15}{l}}
\br
Phase&Inner Radius (mm)&Number of Tapes&Critical Current (A)\\
\mr
1&18&27&127\\
\mr
2&19.6&29&121\\
\mr
3&21.7&31&116\\
\br
\end{tabular}
\end{table}

A study of the first Russian triaxial HTS cable prototype was published in 2017 by the Russian Scientific Research and Development Cable Institute also known as VNIIKP, in collaboration with the 2G-HTS manufacturing company SuperOx~\cite{Fetisov2017IEEE}. This paper produced two prototypes; one being a $10$~m 1G Bi-based HTS cable and the other being a $4$~m ReBCO based 2G-HTS cable. \red{The VNIIKP team inspected the AC-losses at 50 Hz of the 4 m cable with respect to each phase/layer, leading to some results which needed to be investigated from the computational point of view}. In fact, it was made evident how the recording of the AC-losses in the first and second phase, i.e., the inner phases in the triaxial design, could not be trusted due to an unforeseen technical failure in the electrical method used at these phases. Nevertheless, it was agreed that the reported measurements for the AC-losses at the third phase (outer-phase in the triaxial cable) should not have been affected, and therefore it is the purpose of this paper to validate or disregard such observation, \red{starting with a two-dimensional (2D) numerical approach. In this sense, it is to be mentioned that although the 2D approach does not allow to have a faithful representation of the twist pitch for each one of the cable phases, by knowing the experimentally measured critical current at each one of the phases for the prototyped cable, these can be assumed to intrinsically account for the effect of the magnetic mutual inductance of the other phases. Therefore, global quantities such as the magnetization or AC losses of the cable should be reproduced by considering the real critical current density measured at each one of the phases of the triaxial cable (see Table~\ref{Table_1}), these instead of the self-field critical current of a single planar tape that is commonly defined in the classical 2D-modelling. Still, a completely accurate measuring of the magnetic field in the space surrounding the SC tapes cannot be ensured, but as this also depends on the mesh quality invoked, even the 3D models could render to slight deviations from the experimental results. Therefore, a 2D modelling supported by the experimental measurements for the critical current density in the real geometry of the cable for which the AC losses are to be determined, is considered by us as the most sensible approach allowing to avoid the high computational costs associated with the 3D modelling.}

Thus, in this paper we present the numerical estimation of the AC-losses produced by the first Russian triaxial $4$~m ReBCO cable prototype, by the means of a detailed 2D model based on the so-called H-formulation previously used by our group and other researchers in other geometries~\cite{Brambilla2007,Zermeno2011,Ruiz2018SciRep,Ruiz2019MDPI,Ruiz2019JAP,Ruiz2019IOP,Ruiz2019IEEE}. The triaxial cable contains 87 SuperOx $4$~mm width tapes \red{with all their physical properties and dimensions being reported in~\cite{Zhang2018,Lee2014}}. This is done with particular focus on the magnetic profiles of the three phase system and a further interest in the AC-losses within the different phases of the model. From the research conducted, there is an interest in the magnetic interaction between the three superconducting layers (phases), with particular focus on the events occurred at the current peaks such that the correct hysteretic cycle is determined. Our numerical simulations have yield to a \red{good agreement} between our AC-loss calculations and those measured in the triaxial cable, proving the reproducibility and reliability of our method, as well as of the low hysteretic losses produced by the SuperOx triaxial cable by means of its characteristic feature of no magnetic leakage.



\section{Geometry of the triaxial cable and H-Formulation method}\label{Sec.2}

This model features what in the computational jargon for electromagnetic modelling is called as the ``air'' domain, i.e., a no magnetic highly resistive media, it surrounding the cross-section of the triaxial cable which is composed by three concentric rings of finite superconducting tapes, each \red{defining} a phase of the conductor. With the first or inner phase radius being $18$~mm forming a ``ring'' made of 27 tapes, the number of tapes at each phase is then increased by 2 tapes within a no-symmetric arrangement between their radii. The values of the radii and number of tapes are stated in table~\ref{Table_1}, together with the average intensity of the critical current \red{experimentally} measured at each one of the tapes per phase~\cite{Fetisov2017IEEE}. Each tape is composed by a superconducting layer of $4$~mm width and 1~$\mu$m thickness with the critical current density dependence on the magnetic field and angle, $J_{c} (B,\theta)$, as reported in Ref.\cite{Zhang2018}.

\begin{figure}[!]
\centering
\resizebox{1\textwidth}{!}{\includegraphics{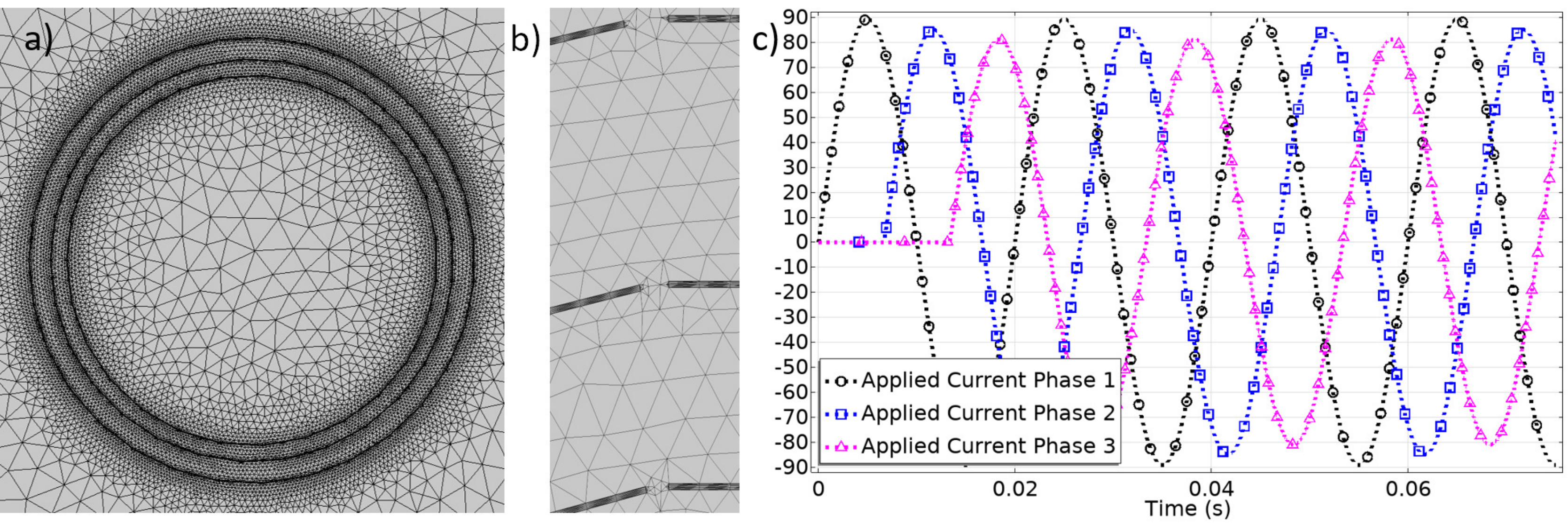}}
\caption{\label{Fig_1} \red{a) 2D representation of the meshed cross-section for the ReBCO triaxial cable prototyped in~\cite{Fetisov2017IEEE}. b) A zoomed-in of the meshing and relative distance between neighbouring tapes in the triaxial arrangement. c)  The unbalanced AC distribution of transport current applied to the three phases of the cable as informed by the experiments and computational needs.}}
\end{figure}

\red{In Figure~\ref{Fig_1}, it is to be noted that the thickness of the superconducting layer within the SuperOx tape has been increased by a factor of 50. This allows to overcome the numerical difficulties that can be caused by the large aspect ratio of the 2G-HTS tapes, whilst maintaining the integrability of the physical quantities by renormalizing the critical current density $J_{c}$ as it is customarily done in the numerical modelling of 2G-HTS tapes \cite{Stenvall2013SUST ,Ruiz2019MDPI, Ruiz2018SciRep}. Likewise, as none of the other layers composing the SuperOx tapes are known to be magnetic nor carry any of the transport current which is to flow only on the superconducting layer, these can be assumed to have the same electrically insulating and non-magnetic properties of the so-called ``air'' domain, what allows to neglect also the occurrence of eddy currents. Same assumptions apply to the electrically insulated stainless steel cable former, as no magnetization signal nor eddy currents have been reported in this case~\cite{Fetisov2017IEEE}.} The varied distance between each phase can be seen in Fig.~\ref{Fig_1}~a) and Fig.~\ref{Fig_1}~b), which is used to provide a sufficient accuracy for the simulations. Fig.~\ref{Fig_1}~c) shows the three-phase unbalanced applied current with phase 2 and 3 being applied from a third of a cycle and two thirds of a cycle, respectively. This is \red{to have a zero-field solution at the initial time instant $(t=0s)$, enabling the time-dependent solution and convergence of the PDE system. This implies to perform the numerical computation for at least 3.75 cycles of the applied current to the first phase (see Fig.~\ref{Fig_1}~c)), with the AC losses for the entire cable being calculated within 0.05166 s and 0.07166 seconds, i.e., being this the first encountered hysteretic cycle after a magnetic relaxation period of 0.05166 s.}

Using the H-formulation in a two-dimensional setting implies to have only two dependent variables, which are the magnetic field in both the \textit{x} and the \textit{y} direction. This two-dimensional arrangement implies that the current density flows only perpendicularly to the magnetic field, i.e., along the \textit{z}-axis. These factors are implemented into COMSOL using the general Partial Differential Equations (PDEs) module, with the magnetic field and the current density being calculated through Faraday's law:
\begin{eqnarray}\label{Eq_1}
\begin{vmatrix} \partial_{y} E_{z}\\ 
-\partial_{x} E_{z} \\ 
\end{vmatrix} 
= - \mu
\begin{vmatrix} \partial_{t} H_{x}\\ 
\partial_{t} H_{y} \\ 
\end{vmatrix}  \, ,
\end{eqnarray}%
with the Ampere's law implemented as a constitutive equation for
\begin{eqnarray}\label{Eq_2}
J_{z} = \partial_{x} H_{y} - \partial_{y} H_{x} 
\end{eqnarray}

The electrical behaviour of the superconducting material is defined using the conventional $E-J$ power law model~\cite{Ruiz2019MDPI}, 
\begin{eqnarray}\label{Eq_3}
\boldsymbol{E} = 
E_0 
\frac {\boldsymbol{J_z}} {\begin{vmatrix}\boldsymbol{J_z} \end{vmatrix}}
\begin{pmatrix} 
\frac {\begin{vmatrix}\boldsymbol{J_z} \end{vmatrix}}
{J_c}
\end{pmatrix}
^{n} \, ,  
\end{eqnarray}
where the standard electric field criterion of 1~$\mu$V/cm has been is adopted. An n-value of 34.4 is used according to the experimental measurements stated earlier for the critical current density dependence $J_{c}(B,\theta)$ for the $4$~mm SuperOx tapes \cite{Zhang2018}.

Special interest has been paid onto the hysteretic AC-losses of the superconducting tapes, which are calculated by the integration of the local density of power dissipation (\textbf{E} $\cdot$ \textbf{J}) across the superconducting tapes (S), and then over the time span for a full hysteretic cycle (\textit{f.c}), i.e,
\begin{eqnarray}\label{Eq_3}
Q = \omega \int_{f.c} dt \int_S \boldsymbol{E} \cdot \boldsymbol{J} dS \, .
\end{eqnarray}
However, in the case of a triaxial cable with an apparently unbalanced distribution of transport current (see Fig.~\ref{Fig_1}~c)), the correct time span for defining a fully hysteretic behaviour has to be computationally determined. \red{The transport current applied to each one of the phases is included within the COMSOL platform as pointwise constraints for each one of the tapes, satisfying that the integral function for the variable $J_{z}$ (Eq.~\ref{Eq_1}) over the cross section of the superconducting domains (per phase) minus the applied current per phase equals zero.}

Finally, it is to remind that the rescaling of the superconducting layer thickness from $1\mu$m to $50\mu$m, and consequently the critical current density, is not only for tackling the large aspect ratio problem of the superconducting tape cross-section, but to include a sufficiently fine mesh across the thickness where the profiles of current density in a could flow \red{within the Bean's approach}. As finer is the mesh as accurate is the motion of the flux front profile inside the superconductor, and therefore as accurate becomes the calculation of the AC-losses. However, as finer is the mesh as higher is the random access memory (RAM) and computational time demanded. In this regard, the numerical calculations shown at \red{Figs.~\ref{Fig_2} \&} \ref{Fig_3} are for a mesh of seven rectangular elements (segments) across the thickness of each one of the 87 superconducting tapes, whose width has been split into a free-triangular mesh with about 28 elements across each one these segments, i.e., each superconducting layer being defined by approximately 196 finite elements \red{within the most general Lagrange shape function with a quadratic element order, this to avoid mathematical reductions that could lead to infeasible physical solutions~\cite{FEM_Bathe_2006}}. The total number of finite elements (1012516), vertex elements (1772), and boundary elements (18821) at each one of our models has been reduced \red{by removing mesh control entities, which are for instance, unnecessary edges within the material domains which are automatically created by the CAD assembly~\cite{FEM_Bathe_2006}. Therefore,} consideration of only 352 vertex elements and 3703 boundary elements, with a minimum element quality of 0.016, led to affordable computing times ranging from 2.7 hours at $I/I_{c}=0.1$ and 3.3 hours at $I/I_{c}=1$ per cycle. A standard PC with a clock-speed per core-processor of 3.6~GHz and 64~GB RAM has been used, with only one core active per computation.  


\section{Results}\label{Sec.3}

\begin{figure}[h]
\centering
\resizebox{0.75\textwidth}{!}{\includegraphics{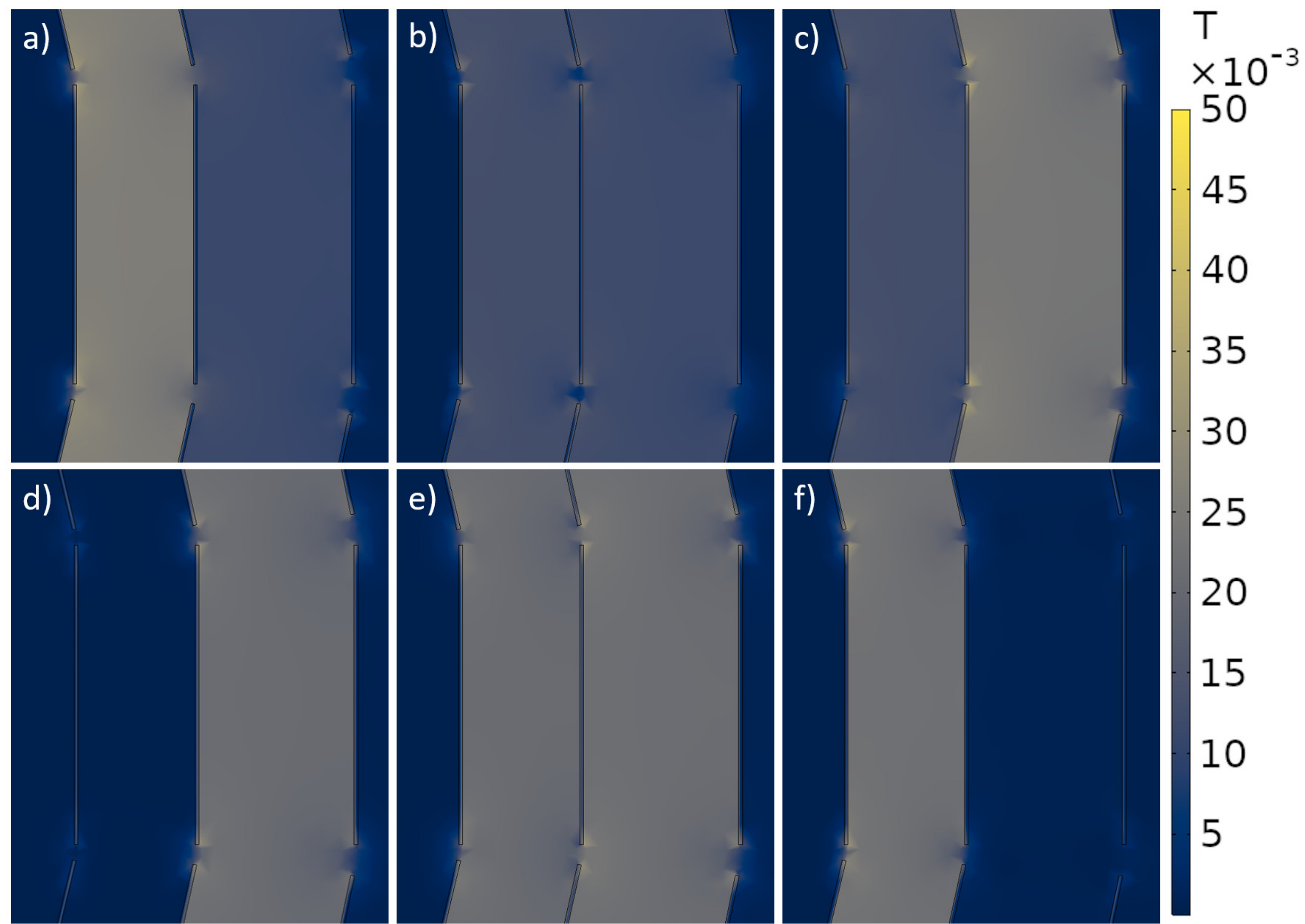}}
\caption{\label{Fig_2} {The magnetic profiles of the cable across a hysteretic time period with (a), (b) and  (c) being when phase 1, 2 and 3 are individually at the peak of the respective phases applied current. The bottom pane of (d), (e) and (f) are when the applied currents of phase 1, 2 and 3 equal 0 A, respectively.}
}
\end{figure}

To analyse the magnetic interaction between the superconducting tapes and the three none equally distanced phases, particular attention has been played towards the magnetic field produced between each phase at both the peak current value and when the applied current is equal to $0$~A. \red{This is due to these values showing both extremes of the electromagnetic response produced by the triaxial cable, that being when either the maximum or zero transport current condition is being applied.}. Therefore, In Fig.~\ref{Fig_2}, the hysteretic magnetic $B-$field distribution for 6 cases are displayed. At the top, each phase reaches its peak applied current, starting from phase 1 \red{(Fig.~\ref{Fig_2}~a))} to phase 3 \red{(Fig.~\ref{Fig_2}~c))}, respectively. The bottom pane shows the corresponding $B-$field profiles when each phase current equals $0$~A. All subplots show no field leakage outside the triaxial cable proving the correct inductive balancing of the same. 

In \red{Fig.~\ref{Fig_2}~a), i.e.,} for the 1st phase peak current of $127$~A, the intensity of the magnetic field $B$ between phase 1 and 2 is (in average), about twice the $B-$field between phase 2 and 3, with the maximum field produced being about 45 mT. Then, \red{as it can be seen in Fig.~\ref{Fig_2}~b),} i.e., for the following peak value of the 2nd phase peak current of $121$~A, a relatively even balance for the $B-$fields between phases is seen. Finally,  in \red{Fig.~\ref{Fig_2}~c), i.e.,} for the 3rd phase peak current of $116$~A, it can be seen an opposite behaviour to what was observed in Fig.~\ref{Fig_2}~a), but showing a slightly lower maximum-field of 40 mT caused by the imbalance in the applied current and the different spacing between the phases. 

Similar response is shown in the bottom pane of \red{Fig.~\ref{Fig_2}~d) to f)} when each phase reaches its zero transport current condition ($I_{tr} = 0$~A), with the other phases at $I_{tr} \neq 0$~A for $\phi = \pm 120\degree$ the phase reference. Thus, when no current is in phase 1 \red{(Fig.~\ref{Fig_2}~d),} nearly no magnetic field is seen in between phases 1 and 2 but a very small field leakage at phase 2 is observed around the tapes. On the other hand, when there is no transport current in phase 2 \red{(Fig.~\ref{Fig_2}~e))}, nearly no screening of the magnetic field is provided by this phase, but the compensation of the field-phase outside the cable demonstrates its current balancing \red{by the continuous absence of magnetic leakage, which agrees with the original optimization method used at~\cite{Fetisov2017IEEE} for the triaxial cable designing.} This is also confirmed \red{by Fig.~\ref{Fig_2}~f),} where no B-field is observed around the phase 3 at the instant of time when only this phase has null transport current.

\begin{figure}[h]
\centering
\resizebox{0.9\textwidth}{!}{\includegraphics{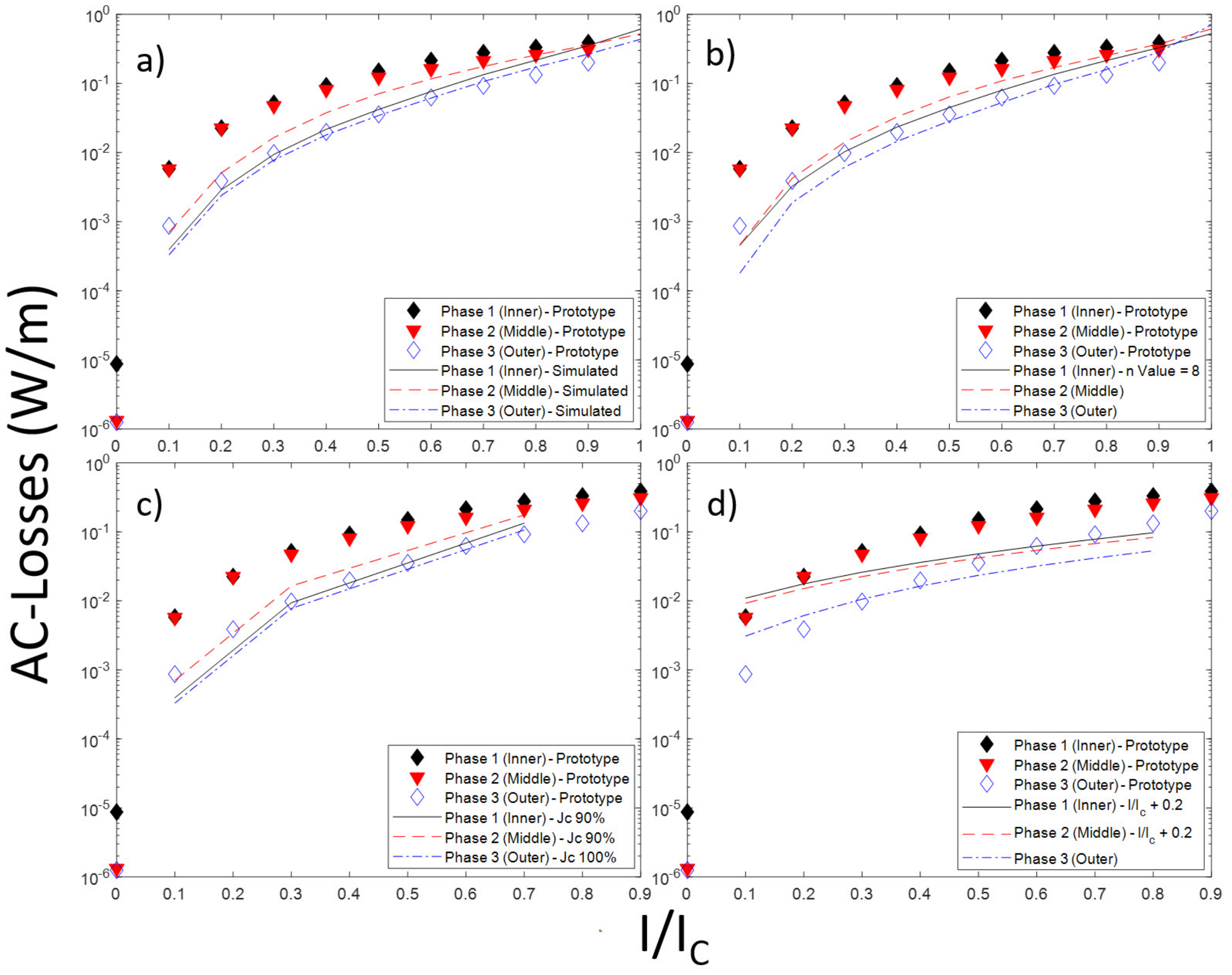}}
\caption{\label{Fig_3} {Calculated AC-Losses with (a) default values, (b) reduced n-values, (c) reduced Jc in phase 1 and 2, and (d) increased transport current by $0.2I_c$ for phase 1 and 2. }
}
\end{figure}

\red{On the other hand, initially the AC-losses caused a concern due to the difficulties in replicating the losses produced within the phase 1 and 2 of the cable. When looking at Fig.~\ref{Fig_3}~a), the calculated AC losses for third phase provides an adequate explanation of the experimental results, but it shows an acute difference at low transport currents for the phases 2 and 3, which both show a very similar but unexpected high losses behavior in the experimental measurements for the applied current range of $0.1I_{c}$ to $0.6I_{c}$, whilst the numerical estimation for currents higher than $0.8I_{c}$ show a good resemblance with the experimental observations. Therefore, despite a proper current balancing of the system and no magnetic leakage was previously demonstrated, it resulted necessary to investigate what could be the potential issues that may have caused these differences with the experimental observations, before concluding the occurrence of technical issues in the experimental setup which were duly discussed with the VNIIKP team. Thus, from the computational perspective, the first attempt was to reduce the n-value within phase 1 from 34.4 down to 8 to see if this would have a drastic change within the AC-losses (see Fig.~\ref{Fig_3}~b)). With phase 2 being the highest AC-losses for most of the applied current range at the default values, by reducing the n-value, this caused the relationship between phase 1 and 2 to become closer at lower applied currents. Unfortunately, it also reduced the AC-losses within phase 2 very slightly, instead of increasing them to justify the large difference. Consequently, degrading the n-value of the superconducting tapes cannot be the reason for the difference between the computationally assessed and measured AC losses for the inner phases of the triaxial cable.}

\red{A second theory was tested to see if there was any significant degradation of the 2G-HTS tapes used in the inner phases, which implied to keep the same applied current for each one of the tapes, but reducing the critical current density of phases 1 and 2 (see Fig.~\ref{Fig_3}~c)). The aim here was to maintain the AC-losses of phase 3 so that they continue to reproduce the experimental observations, whilst finding an alternative critical current density which could increase the AC-losses of phases 1 and 2. The values of $0.1I_{c}$, $0.3I_{c}$ and $0.7I_{c}$ were used as reference points so that we could compare where the largest discrepancy was for the default values. Again, these results did not provide a sound justification for the difference in the AC-losses.}

\red{The final effort to attempt explaining the differences in the AC-losses predicted for the inner phases, was to increase the applied current by an extra $0.2I_{c}$ within phase 1 and 2, assuming a possible but unexpected raise in the amount of current driven by the power sources (see Fig.~\ref{Fig_3}~d)). In this sense, although increasing the applied current of phase 1 and 2 for low values of the stated applied currents ($0.1I_{c}$ to $0.3I_{c}$ with phase 3 as the reference) raised the estimated values for the AC losses within these phases, being now closer with the experimental results, it has the opposite behavior for the phase 3. Moreover, the discrepancies with the experimental measurements become greater as the transport current at each one of the phases increases, what allow us to conclude that the issue encountered is not related with the correct definition of the applied currents shown in Fig.~\ref{Fig_1}~c). Thus, as any degradation of the $J_{c}$ properties of the superconducting tapes nor a technical failure on the power sources can explain the acute raise in the AC losses observed for the inner phases, it has been concluded that the most likely reason for having the significant increase in the AC losses for the phases 1 and 2, with the phase 3 being appropriately reproduced, lies in a possible failure of the electric contacts for the voltage taps fixed at the phases 1 and 2, which could not be revised after the full assembly of the cable, as these were already under the electrical setup used for the outer layer, i.e., the phase 3 of the cable.}

For future publications, we have been allowed to disclose that the VNIIKP-SuperOx team is currently preparing a fully integrated calorimetric setup, aiming to overcome the technical difficulties encountered with the measurement of the AC losses at the inner (first) and middle (second) phase of the triaxial cables. \red{Meanwhile, our team is preparing a comprehensive study on further prototypes currently being developed at the VNIIKP facilities~\cite{Fetisov2021IEEE}.}


\section{Conclusion}~\label{Sec.4}

In this paper, we have focused on \red{the computational modelling} of the first Russian triaxial cable produced by VNIIKP and SuperOx~\cite{Fetisov2017IEEE}. \red{After investigating different scenarios for the forecasting of the AC-losses from the finite elements modelling of the triaxial cable, this in comparison with the attained experimental results from the electrical probe method at each one of its phases, it has been determined that the actual AC losses at the inner phases, i.e., the phase 1 and 2 of the cable are actually lower than the originally reported. In this sense, to support this conclusion we have performed a comprehensive study on the different factors that can alter the critical current density in the different phases of the cable, assuming some degradation of its electrical properties caused by the cable manufacturing, it despite the fact that the critical current at each one of the phases/tapes was correctly measured after the cable assembly. Thus, we have concluded that no modification of the n- and $J_{c}$ values in the $E-J$ power law of the ReBCO conductors could explain the unusual rise in the AC losses observed for the phase 1 and 2 of the cable, nor a failure in the calibration of the power sources. On the contrary, as a correct estimation of the AC losses has been obtained for the outer phase of the cable, i.e., its third phase, and also, it has been determined that the intensity and time dependence of the applied currents with the experimentally measured critical currents after the cable assembly, correctly leads to the proper power balancing of the cable by proving the absence of magnetic leakages, it has resulted clear for our teams that the source of the additional losses in the inner phases is not caused by the cable itself but by the voltage taps setups used for the measuring of the AC-losses in the inner two phases, which cannot be accessed after the cable assembly. In this sense, VNIIKP is currently preparing a fully integrated calorimetric setup which by default, will overcome the technical difficulties encountered with the electrical method for the measurement of the AC losses.} 

Alongside predicting the actual AC-losses produced by the VNIIKP-SuperOx triaxial cable, within the experimental specifications for the cable testing, we have demonstrated that an adequate magnetic field balancing of the system can be achieved, it produced by the non-isotropic arrangement of the phases that has been considered by VNIIKP. At the phase 1 peak current, these results show that the intensity of the magnetic field \textbf{B} between phase 1 and 2 is, about twice the B-field between phase 2 and 3. This behaviour was also observed when phase 3 reached its peak current, but with a slightly lower maximum-field between phase 2 and 3. When the 2nd phase reaches its peak, it produces a relatively even magnetic balance between the inner and outer phases. \red{Likewise, at the instants of time when no transport current flows across anyone of the three phases of the triaxial cable, it has been demonstrated that no magnetic leakages are to be observed even when no magnetic screening of the secondary phase is achieved. This demonstrates the adequate current balancing on the system of transport currents considered by VNIIKP in the Russia's first commercial prototype for triaxial cables.}

%
\section*{References}
\bibliographystyle{iopart-num}
\bibliography{References_Ruiz_Group}

\section*{Acknowledgements} This work was supported by the UK Research and Innovation, Engineering and Physical Sciences Research Council (EPSRC), through the grant Ref. EP/S025707/1 led by H.S.R. All authors acknowledge the use of the High Performance Computing facility ALICE at the University of Leicester. Special thanks are given to Vitaly Vysotsky, Sergey Fetisov, and Vasily Zubko from VNIIKP, for the fruitful discussions and data sharing on their superconducting triaxial cables.

\end{document}